\documentclass[twocolumn]{IEEEtran} 
\IEEEoverridecommandlockouts
\usepackage{hyperref}
\usepackage{graphicx}
\usepackage{filecontents}
\usepackage[noadjust]{cite}
\usepackage{soul}
\usepackage{siunitx}
\usepackage{bm}
\usepackage{adjustbox}
\usepackage{graphicx}
\usepackage{subcaption}
\usepackage{caption}
\usepackage{hhline}
\usepackage{enumitem}
\usepackage{amsmath}
\usepackage{amsfonts}
\usepackage{amssymb}
\usepackage{bm}
\usepackage{multirow}
\usepackage{blindtext, xcolor}
\usepackage{lineno}
\usepackage[suffix=]{epstopdf}
\usepackage{listings}

\begin{document}

\title{Extending Pretrained Segmentation Networks\\
with Additional Anatomical Structures}

\author{\IEEEauthorblockN{Firat Ozdemir,
Orcun Goksel}\\
\IEEEauthorblockA{Computer-Assisted Applications in Medicine (CAiM) \\ ETH Zurich, Switzerland}}

\maketitle
\thispagestyle{plain}
\pagestyle{plain}

\begin{abstract}
For comprehensive surgical planning with sophisticated patient-specific models, all relevant anatomical structures need to be segmented.
This could be achieved using deep neural networks given sufficiently many annotated samples, however datasets of multiple annotated structures are often unavailable in practice and costly to procure.
Therefore, being able to build segmentation models with datasets from different studies and centers in an incremental fashion is highly desirable.
\\
We propose a class-incremental framework for extending a deep segmentation network to new anatomical structures using a minimal incremental annotation set.
Through distilling knowledge from the current network state, we overcome the need for a full retraining. 
\\
We evaluate our methods on 100 MR volumes from SKI10 challenge with varying incremental annotation ratios. 
For 50\% incremental annotations, our proposed method suffers less than 1\% Dice score loss in retaining old-class performance, as opposed to 25\% loss of conventional \emph{finetuning}.
Our framework inherently exploits transferable knowledge from previously trained structures to incremental tasks, demonstrated by results superior even to non-incremental training:
In a single volume one-shot incremental learning setting, our method outperforms vanilla network performance by \textgreater11\% in Dice.
\\
With the presented method, new anatomical structures can be learned while retaining performance for older structures, without a major increase in complexity and memory footprint; hence suitable for lifelong class-incremental learning.
By leveraging information from older examples, a fraction of annotations can be sufficient for incrementally building comprehensive segmentation models. 
With our meta-method, a deep segmentation network is extended with only a minor addition per structure, thus can be applicable also for future network architectures.

\end{abstract}
\section{Introduction}
\label{sec:introduction}

In 2014, there were more than 1\,million hip and knee replacement surgeries in total in the US, which is expected to grow even more up to 171\% by 2030.
Based on the records from 2000 to 2014, revision surgeries are similarly expected to increase more than 142\% by 2030, summing up to almost 200 thousand~\cite{aaos2018revision}.
Patient-specific planning and execution of interventions can help to make the surgeries more precise, reducing revision and correctional interventions.
With the increasing life expectancy, growing population, and the major impact of orthopedic conditions on quality of life, personalized interventions should not burden the already increasing costs.
A major source of cost comes from the resources required to model anatomical structures of a patient (e.g.,\ in order to plan guides for osteotomy, craft/pick the right implants for replacement surgery).
Furthermore, additional musculoskeletal structures (i.e.,\ muscles, tendons, ligaments) which may influence the surgical outcome are often completely ignored in interventional planning.
In pursuit of surgical planning for functional outcomes, accurate segmentation of many anatomical structures of interest is essential.

Convolutional Neural Networks (CNNs) are suitable candidates for surgical planning since their inference is fast and they can handle complex segmentation tasks~\cite{Ronneberger2015unet,baumgartner2017exploration,yang2017suggestive,ozdemir2018active}.
For an ultimate goal of functional planning, all relevant anatomical structures will eventually be needed.
However, conventional CNNs require plenty of data with corresponding annotations, and
manual annotation is costly, especially in the medical field given the need for clinical experts.
Intuitively, for similar tissue types (e.g., different bones) knowledge from one class annotation can be transferred for segmenting another.
Following satisfactory segmentation performance for a given set of anatomical structures (e.g., through CNNs), available resources for manual annotation can then be directed to any remaining anatomical structures of interest.
Furthermore, with a successful knowledge transfer method, a diminishing amount of additional annotations per structure could be sufficient in the long term.

The most straightforward way to extend a pretrained network is the so-called \emph{finetuning}, i.e.\ to continue optimization for a new dataset, with potentially new class labels.
This, however, has been shown to cause \textit{catastrophic forgetting}~\cite{mccloskey1989catastrophic} on the previous dataset/classes.
\textit{Class-incremental learning} has the potential to address this issue, since the goal is to achieve high performance on both new and old class categories, while having a bound on or a relatively slow increase of computational cost and memory footprint~\cite{rebuffi2016icarl,castro2018endtoend,ozdemir2018learn}.
The state-of-the-art approach for class-incremental learning for classification task, iCaRL~\cite{rebuffi2016icarl}, exploits the concept of knowledge distillation~\cite{hinton2015distilling} to retain old-class classification performance. 
\textit{Distillation}~\cite{hinton2015distilling} has been initially introduced for training simpler networks using the output of trained complex-network as ``soft targets''.
Other applications of distillation for preserving old-class classification accuracy have been shown in~\cite{rebuffi2016icarl,li2016learning}.

It is essential in lifelong learning to estimate/bound the expanding needs of computational power and memory footprint.
In~\cite{rebuffi2016icarl}, an approach was proposed to keep a predefined number of total samples, so-called \textit{exemplars}, as a representation of previously observed data.
In a different domain, known as \textit{Active Learning for Segmentation}, querying ``representative'' samples was shown as an alternative for selecting a subset from a larger sample set\cite{yang2017suggestive,ozdemir2018active}. 
For this purpose, a maximum set-coverage~\cite{hochbaum1997approximating} problem was solved iteratively; as an alternative to picking samples that are closest to a predefined number of cluster means in a latent feature space as in~\cite{rebuffi2016icarl}. 
While lifelong learning is not a new concept, to the best of our knowledge, it has not been explored for class-incremental segmentation in the imaging community.

In this work, we propose a meta method as an architectural extension to any given network, in order to enable incremental learning for segmentation. 
Although we conduct our studies on a Unet similar to~\cite{Ronneberger2015unet,baumgartner2017exploration}, our proposed extension is simple and can easily be adapted for most other segmentation architectures. 
We further show the significance of distilling knowledge from the previous state of the network when incrementally adapting to new data.
In addition, we also demonstrate the benefit of keeping informative samples throughout a lifelong learning scenario.
We have presented the preliminary results of this work in~\cite{ozdemir2018learn}.
Our specific contributions herein are:
\begin{enumerate}[noitemsep,topsep=0pt]
\item Analysis on the larger and publicly available SKI10 dataset, targeting the knee anatomy and related surgical interventions.
\item Multiple experimental settings to assess feasibility and generalizability of our proposed framework.
\item Study of incremental set imbalance.
\item Investigating physical image resolution effects on our incremental learning.
\end{enumerate}

Below, we first describe our proposed architectural changes on a typical network to enable class-incremental segmentation. 
Then, we explain how knowledge from previous dataset is retained through distillation. We also propose to increase knowledge distillation through informative sample retention.

\section{Methods}

In most medical imaging modalities, appearance of the same tissue, even if in different anatomical structures (e.g., two different muscles) are similar or related. 
Therefore, training separate networks for datasets of different structures which share similar contextual and/or texture content may be redundant, if not suboptimal. 
Below we describe our proposed method in the context of class-incremental learning. 
Without loss of generality, consider incremental step $i$ when an \textit{incremental dataset} $D_i$ is introduced, for the model to be extended with the knowledge therein.
Given our class-incremental focus, we mainly consider the setting where $D_i$ contains new classes.
The cumulative set of all previous data, i.e.\ $\{D_j | j=[0,..,i$$-$$1]\}$, is then called the \textit{old} dataset.

\vspace{1ex}\noindent
\textbf{Incremental Learning of New Structures.}

\begin{figure}[t]
    \centering
    \includegraphics[
			width = \linewidth
            ]{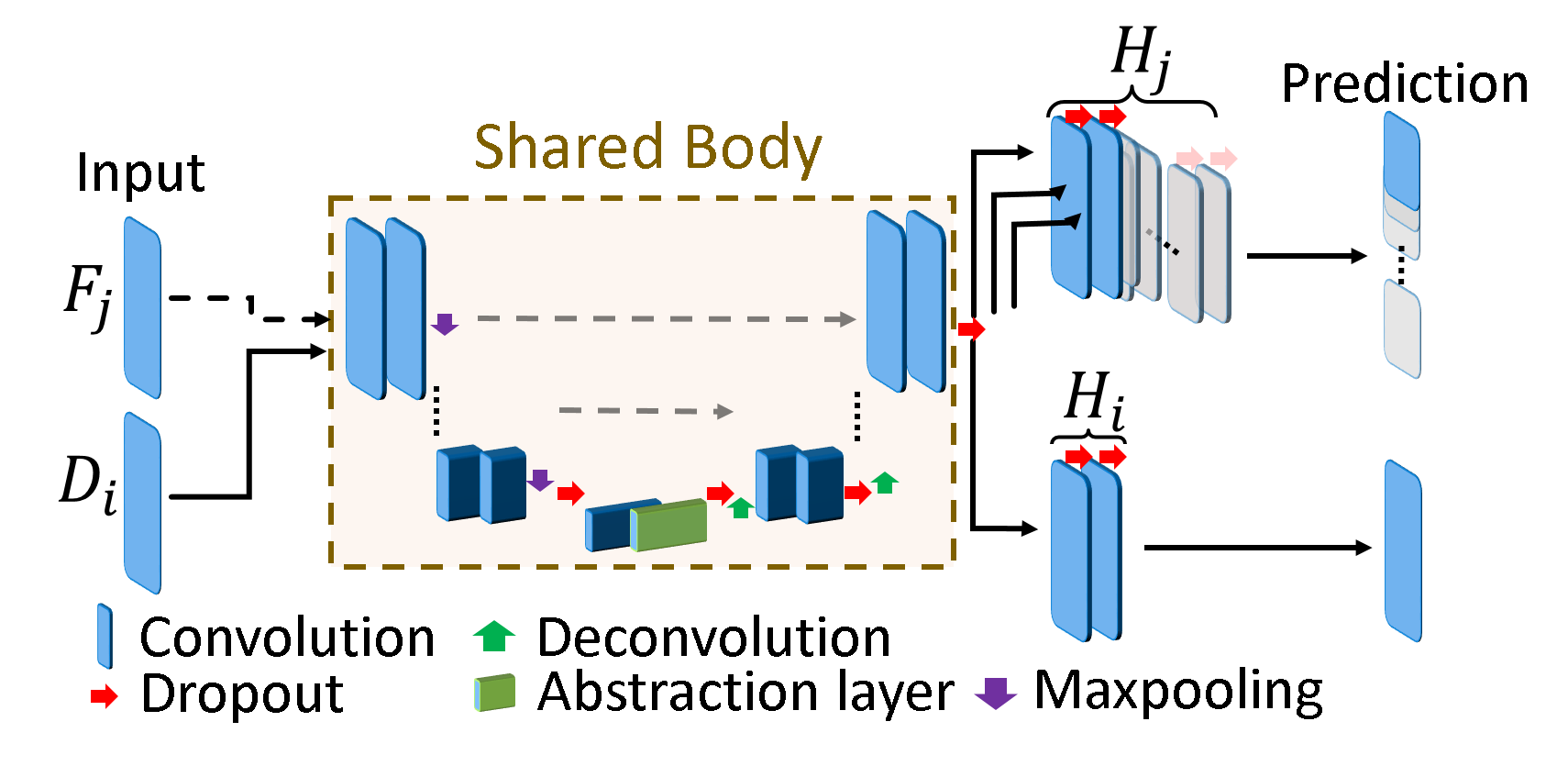}
    \caption{Architecture proposed for lifelong learning for segmentation.
    $D_i$ incremental dataset and its corresponding \textit{head} $H_i$ for predictions, $F_j$ exemplar set (only for AeiSeg) and corresponding heads $H_j$ where $j \in [0,..,i-1]$.} 
    \label{fig:architecture}
\end{figure}

We propose a cost-effective framework as a solution to lifelong learning for class-incremental segmentation.
Our idea is to modify a generic segmentation network (e.g., a U-Net), where an additional block of convolutional layers is appended in parallel at the end of the network \textit{body} as shown in Fig.~\ref{fig:architecture} to incorporate new information at each incremental step.
We call each branching by such additional convolutional layers as a \textit{head} (cf.\ $H$ in Fig.~\ref{fig:architecture}), where a head can be segmenting one or more target structures. 
While such extension is computationally rather inexpensive in terms of the number of additional parameters to be optimized, it turns out to be sufficient to combine and alter the generic features extracted at the network \textit{body} while the corresponding \textit{head} is trained for a specific dataset.

\vspace{1ex}\noindent
\textbf{Retaining Knowledge Through Distillation.}
In the proposed \textit{multi-headed} architecture, the majority of the learned kernels will be shared for all the heads.
In order to prevent \textit{catastrophic forgetting}, one has to ensure that the old heads $\{H_{j} | j=[0,..,i$$-$$1]\}$ can retain their segmentation performance for the classes in the old dataset. 
This can be achieved for the purpose of incremental learning through so-called \textit{soft targets}~\cite{hinton2015distilling}. 
Herein, prior to training with a new dataset $D_i$, we generate soft targets $t_j$ for the existing segmentation labels, as the soft-maxed logits from all old heads $\{H_{j} | j=[0,..,i$$-$$1]\}$ for the new images in $D_i$.
The aim is to take the corresponding soft targets also into consideration for all weights that concern a given head (i.e.,\ shared body and the convolutional blocks in the head) when optimizing the network weights for the incremental dataset. 
Hence, the network is jointly optimized for both segmenting the incremental dataset $D_i$ and producing the same prediction proposals $t_j$ for the old heads.
While any objective function of choice $L_\mathrm{seg}$ can be used for the new head, all old heads are meanwhile optimized for the distillation loss
\begin{equation}
L_\mathrm{dis} = \sum_j \sum_c w_j^{(c)} t_j^{(c)} \log\,p_{j}^{(c)}\, \mathrm{,}
\vspace{-2ex}  
\end{equation} 
where $j \in \{0,..,i-1\}$ is an identifier for the old heads, $t_j^{(c)}$ is the soft target for class $c$ in head $j$, $w_j^{(c)}$ is the desired weighting scalar for the corresponding class and head, and $p_{j}^{(c)}$ is the prediction proposal of head $j$ for class $c$ during the incremental training.
Total loss then becomes $L_\mathrm{total} = L_\mathrm{seg} + L_\mathrm{dis}$, which is defined for the training stage as an aggregate over all pixels normalized by the number of pixels.
We call this approach \textbf{LwfSeg}, a naming inspired by Learning without Forgetting~\cite{li2016learning}.
LwfSeg is a viable option when sharing patient image data for lifelong learning may be problematic due to ethic and privacy concerns, whereas a trained model could be shared with ease.

\vspace{1ex}\noindent
\textbf{Keeping Informative Samples}
In an incremental learning setting, in addition to distillation loss, keeping samples from the original old datasets could greatly help anchor the incrementally learned models to original data.
This becomes of ever growing interest, as more and more medical image datasets are being made publicly available while the numbers are expected to grow further. 
One can, for instance, utilize such datasets for transfer learning by first pretraining models with these and then \emph{class-incrementing} them with an own proprietary dataset, e.g.\ as in~\cite{ozdemir2018learn}.
For retaining knowledge from such a pretraining dataset, the most intuitive approach would be to simply keep all the samples from the original dataset to use these for an $L_\mathrm{seg}$ loss when optimizing the corresponding head.
However, such an approach is not scalable and would lead to major limitations both for storage of such ever-growing data and for processing during training to ensure iterations over all such data; aggravating further at each incremental learning stage. 
An \emph{informed} subset of the old dataset would help in focusing limited training resources. 

Ideally, one can try ($i$) to determine samples that are most representative of the entire old dataset; i.e.,\ cluster means of the representations of the dataset distribution (similar to class means in~\cite{rebuffi2016icarl}).
However, given the class imbalance within datasets, it is likely ($ii$) that some of such representative image clusters will be redundant for the segmentation task at hand; e.g. consecutive slices from a 3D volume.
In order to account for both of these conditions (i) and (ii), we propose to first pick a large batch of samples for which the trained model is ``confident''~\cite{ozdemir2018learn}, and then to prune them with the aim of maximizing how well they represent the entire dataset~\cite{yang2017suggestive}. 

\vspace{1ex}\noindent
\textbf{Model Confidence Approximation.}
Estimating model confidence for a CNN prediction has been of major interest in research, with several approaches proposed.
Although conventional CNNs are deterministic during inference by nature, a recent study~\cite{gal2015dropout} proposed a simple way to obtain Monte-Carlo estimates from an architecture through the use of Dropout layers~\cite{srivastava14dropout} at test time.
Using $n_\mathrm{MC}$ prediction samples $\{S_k(I[x],c)\,|\,k$$=$$[1,..,n_\mathrm{MC}]\}$ for a given class $c$ and a flattened image $I$$\in$$\mathbb{R}^Z$ with $Z$ being the number of pixels, we herein estimate the model confidence $m$ for a prediction using the average pixel variance per Monte-Carlo dropout sample as follows:
\begin{equation}
m(I,c) = 
 \begin{cases}
      -\displaystyle\frac{1}{Z}\sum_{x=1} ^Z \mathrm{var}\Big(\big\{S_k(I[x], c)\}\, 
      \Big)\,, & \text{if class } c \text{ in } I\\
      -\infty\,, & \text{otherwise}\ .
    \end{cases}
\end{equation}
This computes the variance along the dimension $k$ for $S_k(I,c) \in \mathbb{R}^{Z \times n_\mathrm{MC}}$, where the result is averaged spatially over all pixels $x$.
Any image $I$ in which a class of interest $c$ is not annotated is ignored. 

During an incremental step, out of the previous dataset $D_{i-1}$ we pick a confident subset $E_{i-1}$ of images and annotations to preserve, where $E_{i-1}$ contains $n_\mathrm{conf}\cdot|c|$ images $I$ for which $m(I,c)$ is maximized for each class $c$, and $|c|$ is the number of classes in $D_{i-1}$.

\vspace{1ex}\noindent
\textbf{Sample Pruning.}
Within the procured set $E_{i-1}$, there could be many samples which are almost identical (e.g., neighbouring slices from 3D) for which the trained model is equally confident.
Hence, redundant samples can be reduced by dropping samples which have similar looking alternatives. 
Although similarity can be quantified using intensity-based metrics (e.g.\, via difference or correlation) or hand-crafted features, data-driven features from a network trained on a particular datset for a particular task would be expected to better represent discriminative aspects of images relevant to that given task within that given dataset.
Accordingly, similarly to image descriptors in~\cite{yang2017suggestive} we represent an image $I$ as a vector by global average pooling of the feature map (i.e.,\ averaging over the spatial dimensions) at the layer of information bottleneck, i.e.\ the so-called \emph{abstraction} layer seen in Fig.~\ref{fig:architecture}. 
We can then define a distance metric to quantify how similar one image is to another within $D_{i-1}$ by using any similarity metric of choice (e.g.,\ inner product or cosine similarity) between two such vectors.
Given such distance metric, any clustering technique can then be used to find representative images spanning this vector space.
We use maximum \emph{set coverage}~\cite{hochbaum1997approximating} to find a representative set $F_{i-1}\subseteq E_{i-1}$ by iteratively populating $F$ with up to $n_\mathrm{rep}$$\leq$$n_\mathrm{conf}$ images picked from $E_{i-1}$ using the above image descriptor distance.
$F_{i-1}$ is then the set to keep from dataset $D_{i-1}$.
At incremental step $i$, the union of all sets kept from all previous increments can then be indicated as  $\cup_{j=0}^{i-1} F_j$.

During the incremental training, batches are randomly generated from either incremental dataset $D_i$ or the union exemplar set $\cup_{j=0}^{i-1} F_j$.
If the batch consists of exemplars from $F_j$, then the corresponding head $H_j$ and the shared body weights are updated based on the distillation loss $L_\mathrm{dis}$ computed at the prediction of head $H_j$. 
Otherwise, as in LwfSeg, the model weights are updated based on the segmentation loss $L_\mathrm{seg}$ for the new head $H_i$ and the cumulative distillation loss at the older heads, i.e.\ $\sum_{j=0}^{i-1}L_\mathrm{dis}(H_j)$.
We call the above-described extension of LwfSeg by preserving an exemplar dataset as Abstraction-layer Exemplar-based Incremental Segmentation (\emph{AeiSeg}).

\section{Experiments and Results}
\label{sec:experiments}

\noindent
\textbf{Data.}
We have comparatively studied our proposed class-incremental methods on publicly available MR Dataset SKI10 MICCAI Grand Challenge~\cite{heimann2010segmentation}. 
The dataset consists of 100 knee volumes (10874 2D slices) collected at over 80 centers from different vendors with a voxel spacing of $0.4\times0.4\times1.0$\,mm.
Due to varying field-of-view, voxel resolutions of volumes are not consistent in SKI10. 
In order to best utilize available computational power, we resized all in-plane (sagittal) image slices to $224\times224$\,px and used this resolution during training unless stated otherwise.
Reported results in the section below respect the original image resolution of $0.4\times0.4$\,mm, which we achieved through bilinear upsampling of the network logits per label, prior to final segmentation. 
The majority of images are T1-weighted from 1.5\,T MR machines, although images from 1\,T and 3\,T machines as well as some T2-weighted images also exist in SKI10.
Since the imaging settings are not available, we treated all images equally in our experiments. 

\vspace{1ex}\noindent
\textbf{Methods.}
We compared our methods first with two basic non-incremental approaches: \emph{CurSeg} (a single head model trained solely on the previously available dataset; i.e.\ current dataset, $D_{i-1}$) and \emph{IncSeg} (a single head model trained solely on the incremental dataset).
Due to a lack of state-of-the-art, as a naive baseline \emph{finetuning} was also evaluated in the following experiments to serve as a lower bound.
For the finetuning method, to allow a direct comparison with LwfSeg we apply the same multi-headed architecture, but after appending a new head for the incremental data we use only $L_\mathrm{seg}$ as the objective function for the incremental dataset $D_i$.
To serve as a lower-bound for exemplar selection, we create an additional class-incremental learning method with the same architecture; random exemplar-based incremental segmentation (ReSeg).
For each class, ReSeg randomly selects exemplars among the images where corresponding class pixels exist.
Without loss of generality, for extended evaluations we used annotations of femur bone vs.\ background as the \textit{current} (Cur) dataset $D_{i-1}$, while tibia bone vs.\ background as the \textit{incremental} (Inc) dataset $D_{i}$.
To study generalizability per structure order, we also repeated some experiments for the opposite incremental order, i.e.\ Cur being the tibia and Inc being the femur.

\vspace{1ex}\noindent
\textbf{Experimental Scenarios.}
Similarly to a leave-patient-out approach, the dataset was split into \textit{current}, \textit{incremental}, \textit{validation} and \textit{test} sets per patient volume, preventing slices from a volume to exist in more than one set (cf.\ Table~\ref{tbl:experimental_cases}).
In some scenarios many images and annotations may exist for incremental learning, whereas in other settings only a very small set may be available, also referred to as few-shot or one-shot learning.
Ideally, after training with a large dataset, one should need only a small set of new anatomy annotations, leveraging appearance similarities and common information in the field-of-view.
In order to study the effects of incremental dataset size, we experimented with 4 different incremental ratios (IRs) from current to incremental dataset as shown in Table~\ref{tbl:experimental_cases}, where the incremental dataset size is logarithmically scaled.
Given 70 volumes for training, a range of current-to-incremental dataset ratios are then tested, from a balanced set of 35/35 ($=100\%$) in the IR100 experiment to an extremely small incremental set of 1 volume ($1/69=1.44\%\approx1\%$) in the IR01 experiment, i.e.\ one-shot learning.

\subsection{Implementation Details}

The proposed modified UNet architecture is developed for 2D images with single channel, where the number of filters $n_\mathrm{fil}$ at the first convolutional layer is 32.
The rest of the layers follow the same architecture as UNet, where the number of filters are defined as $2^{l}n_\mathrm{fil}$ where $l$ is the level of spatial coarsening.
We set all heads to consist of 2 convolutional layers each with $n_\mathrm{fil}$ filters.
All convolutional layers are batch normalized prior to ReLU activation and consist of $3\times3$ kernel, with the exception of final prediction layer, which has a $1\times1$ convolutional kernel.
Spatial dropout layers with a rate of 0.5 were utilized as shown with red arrows in Fig.~\ref{fig:architecture}.
Note that additional dropout layers were used along the spatial upscaling path prior to deconvolution filters (see supplementary material for details). 
Each experiment was conducted with 4 images per batch.
Every method was trained for a fixed number of steps $n_\mathrm{stp}=20000$ (i.e.,\ for $n_\mathrm{stp}$ batches) using Adam optimizer with a learning rate of $0.001$.
We used inverse-frequency weighted cross-entropy loss for $L_\mathrm{seg}$.
Similarly, the class weights $w$ in $L_\mathrm{dis}$ are computed from the training set of the corresponding dataset as the inverse label frequency.
For the incremental methods (finetuning, LwfSeg, and AeiSeg), an additional training of $n_\mathrm{stp}$ steps was performed, initialized by the best validation set state of CurSeg. 
We have used cosine similarity as the distance metric when populating representative set $F$.
Batches were randomly scaled and horizontally flipped during training for the purpose of data augmentation.
Our TensorFlow implementation is publicly available\footnote{\url{https://github.com/firatozdemir/LwfSeg-AeiSeg}}.

In order to prevent overfitting to training set, model Dice score on a separate validation set was used to determine the state of a model to be used on the test set. 
For the incremental methods, we made an implementation choice to pick the model state which maximizes the average Dice score on current and incremental classes together on the validation set.
However, for the finetuning method, this resulted in using the model state at a training step as early as 100, due to very early forgetting of the current class knowledge.
Comparisons with finetuning with maximum incremental class performance were presented earlier in~\cite{ozdemir2018learn}, indicating poorer baseline results.
For AeiSeg, we empirically picked $n_\mathrm{MC}=29$, $n_\mathrm{conf}=1000$ and $n_\mathrm{rep}=100$.
\begin{table}
\centering
\normalsize
\caption{Data partitioning for experiments. IRXX represents (rounded) ratio XX\% of incremental (Inc) annotated volumes to previously available (Cur).}
\label{tbl:experimental_cases}
\adjustbox{max width=\textwidth}{
\begin{tabular}{r|cccc}
\#Volumes & IR100 & IR17 & IR04 & IR01 \\ \hline
Current (Cur) & 35 & 60 & 67 & 69 \\
Incremental (Inc) & 35 & 10 & 3 & 1 \\
Validation & 5 & 5 & 5 & 5 \\
Test & 25 & 25 & 25 & 25 \\
\end{tabular}
}
\end{table}

\subsection{Results}
\label{sec:results}

\begin{figure*}[]
\centering
	\begin{subfigure}[b]{0.249\textwidth}
    \centering
        \includegraphics[width=0.99\textwidth, trim= 0cm 0.0cm 0cm 0cm, clip]{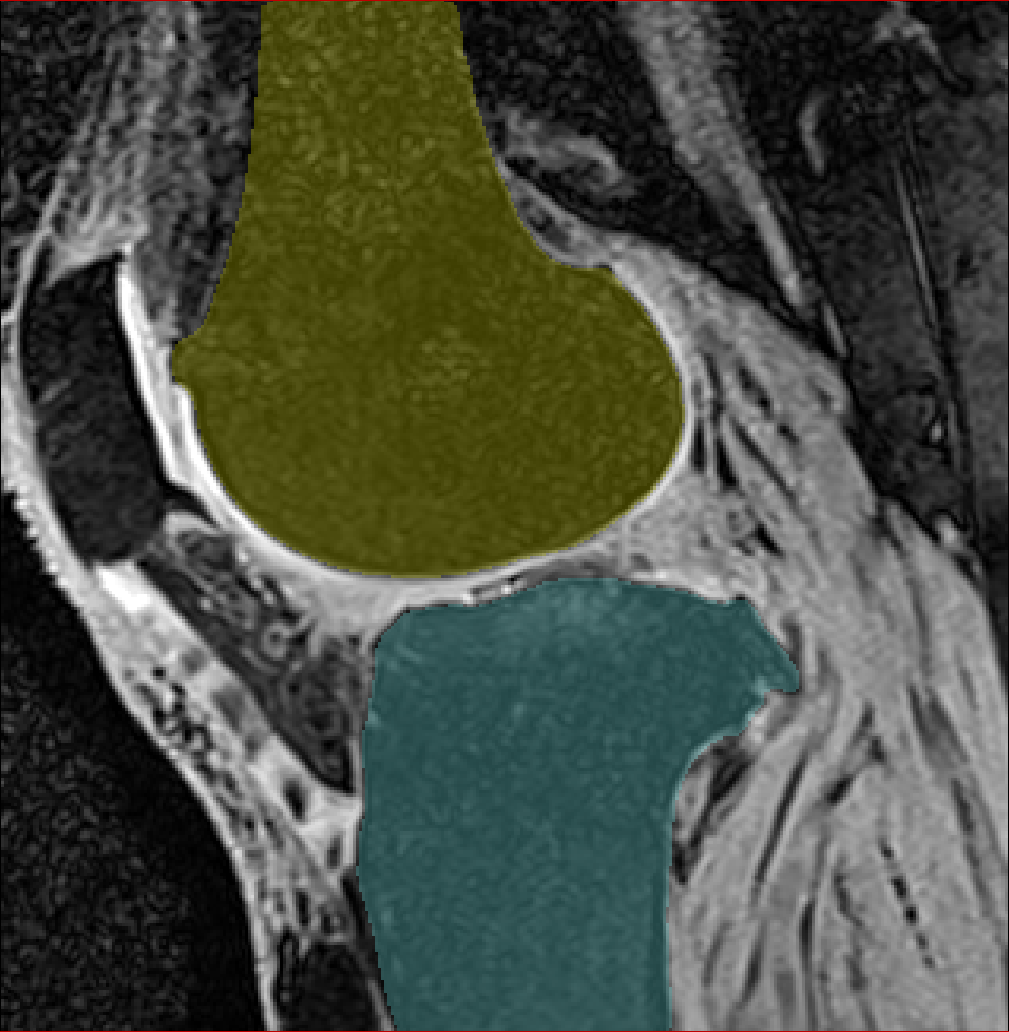}
        \caption{Ground truth}
    \end{subfigure}%
    \begin{subfigure}[b]{0.249\textwidth}
    \centering
        \includegraphics[width=0.99\textwidth, trim= 0cm 0.0cm 0cm 0cm, clip]{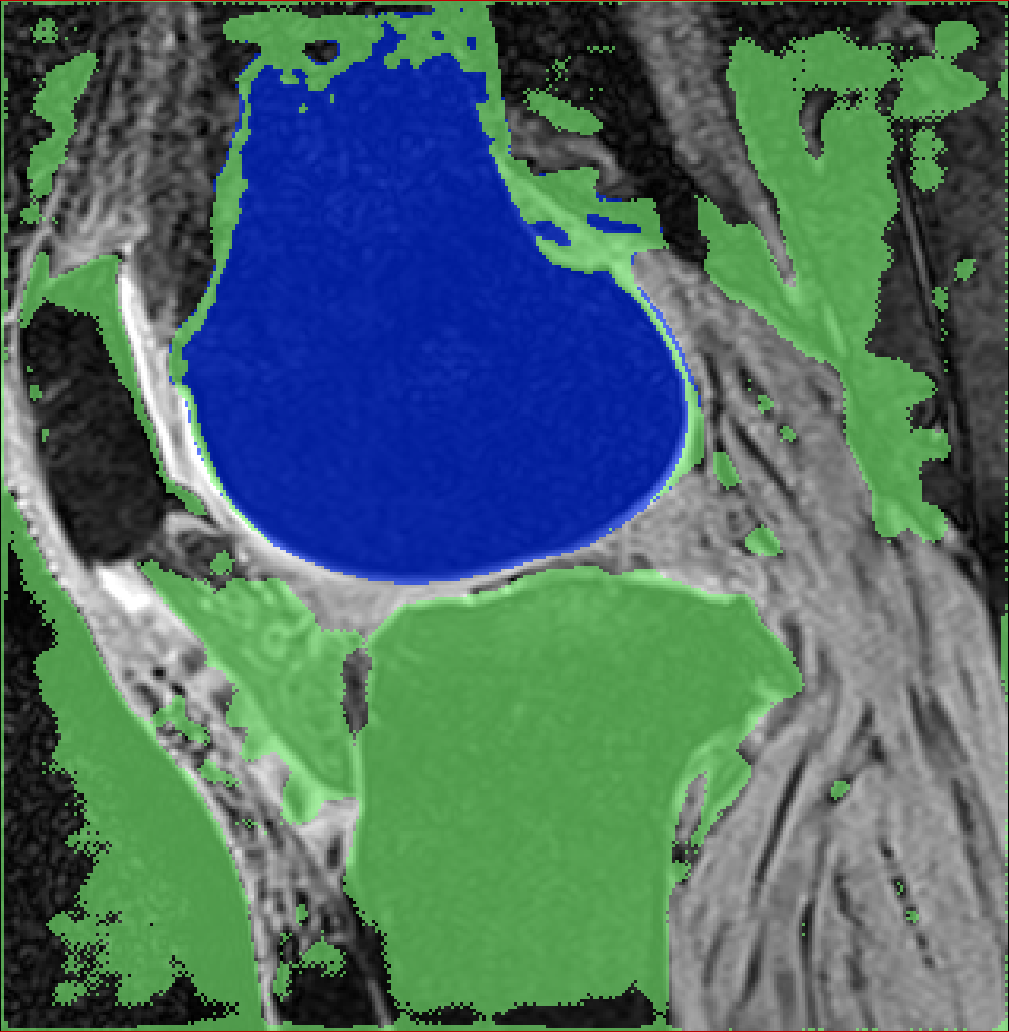}
        \caption{Finetune}
    \end{subfigure}%
    \begin{subfigure}[b]{0.249\textwidth}
    \centering
        \includegraphics[width=0.99\textwidth, trim= 0cm 0.0cm 0cm 0cm, clip]{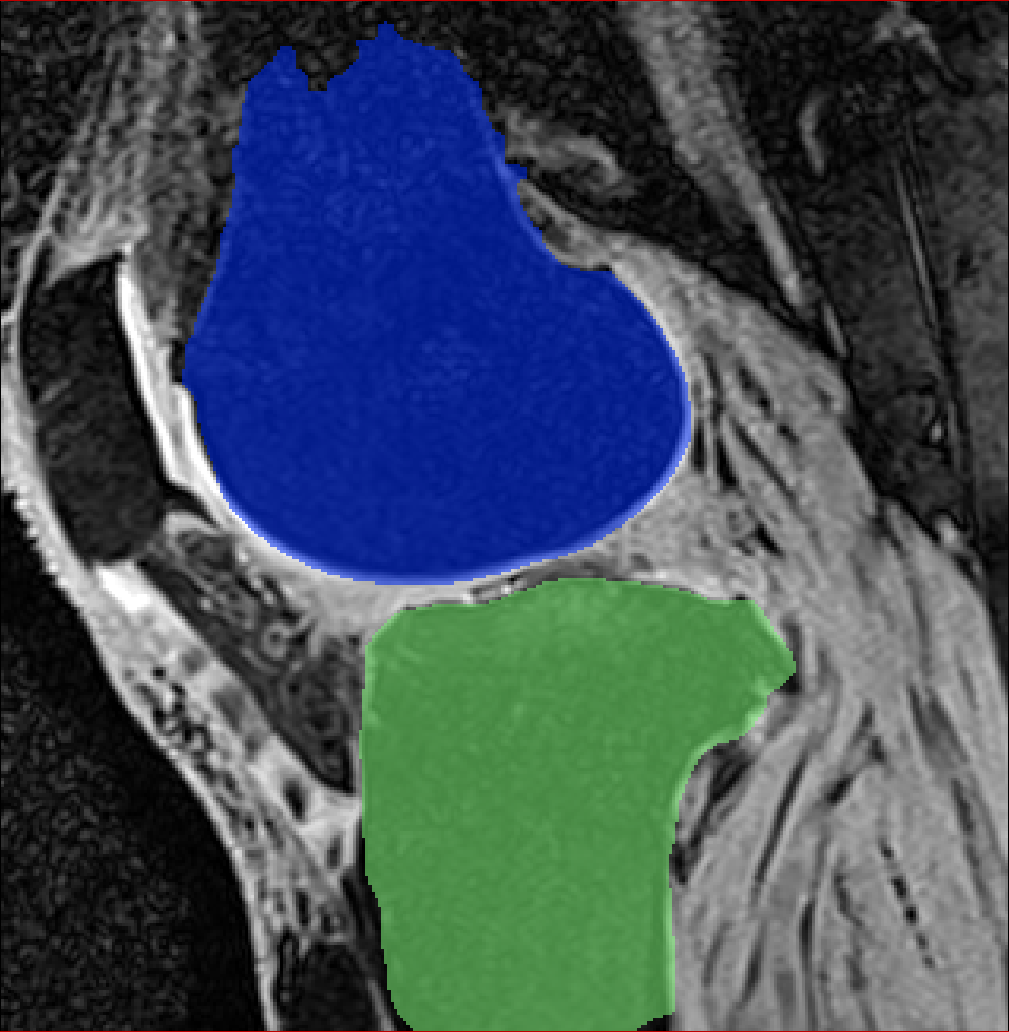}
        \caption{LwfSeg}
    \end{subfigure}%
    \begin{subfigure}[b]{0.249\textwidth}
    \centering
        \includegraphics[width=0.99\textwidth, trim= 0cm 0.0cm 0cm 0cm, clip]{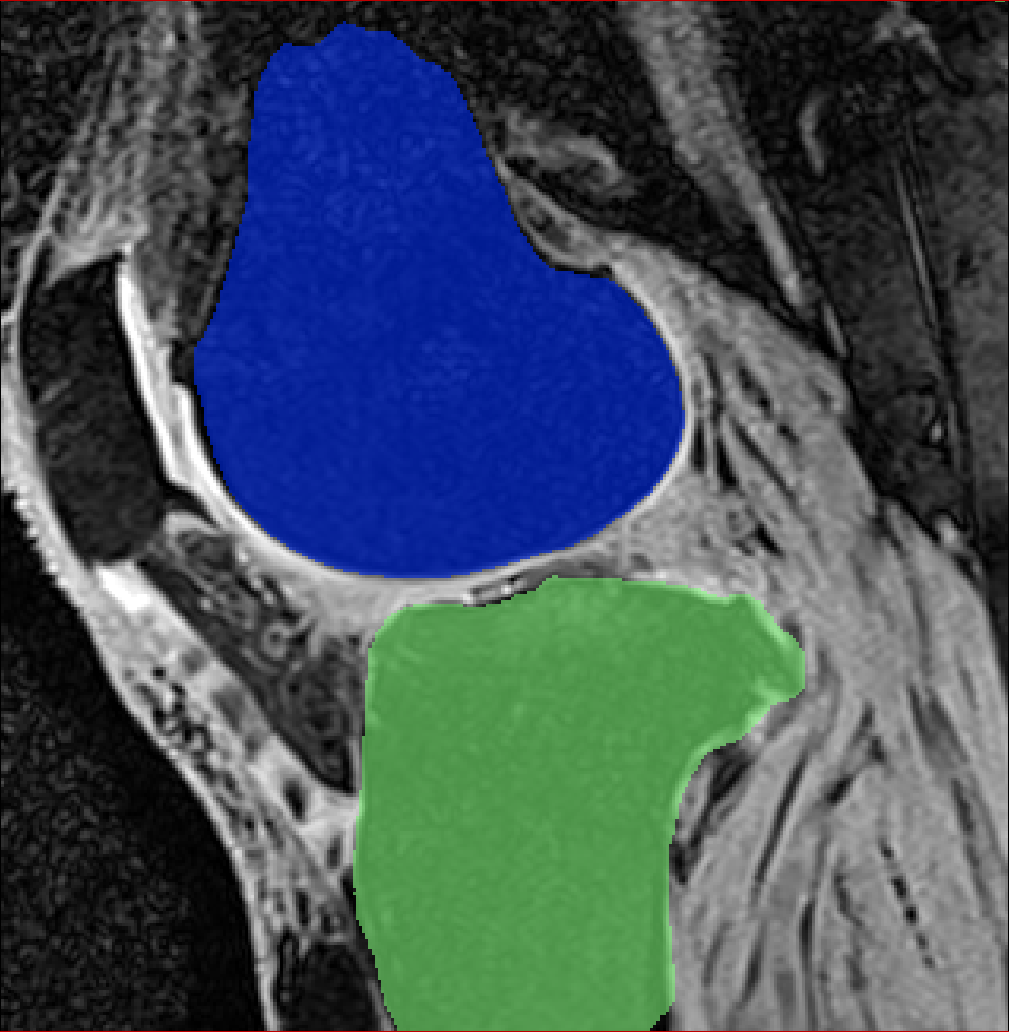}
        \caption{AeiSeg}
    \end{subfigure}%
\caption{An example slice from an IR01 experiment from volume 12 comparing finetune, LwfSeg, and AeiSeg with respect to ground truth annotations.}
    \label{fig:qualitative_comparison}
\end{figure*}

\begin{figure*}[]
\centering
	\begin{subfigure}[b]{0.249\textwidth}
    \centering
        \includegraphics[width=0.99\textwidth, trim= 0cm 0.0cm 0cm 0cm, clip]{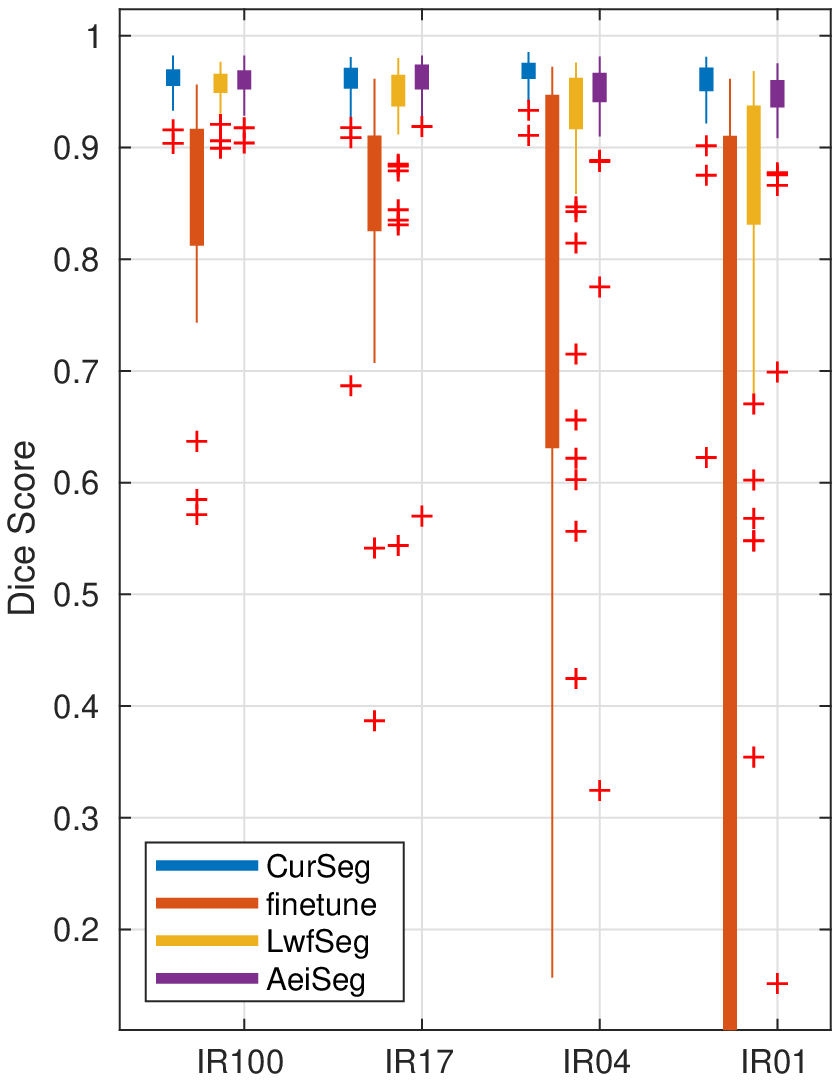}
    \end{subfigure}%
    \begin{subfigure}[b]{0.249\textwidth}
    \centering
        \includegraphics[width=0.99\textwidth, trim= 0cm 0.0cm 0cm 0cm, clip]{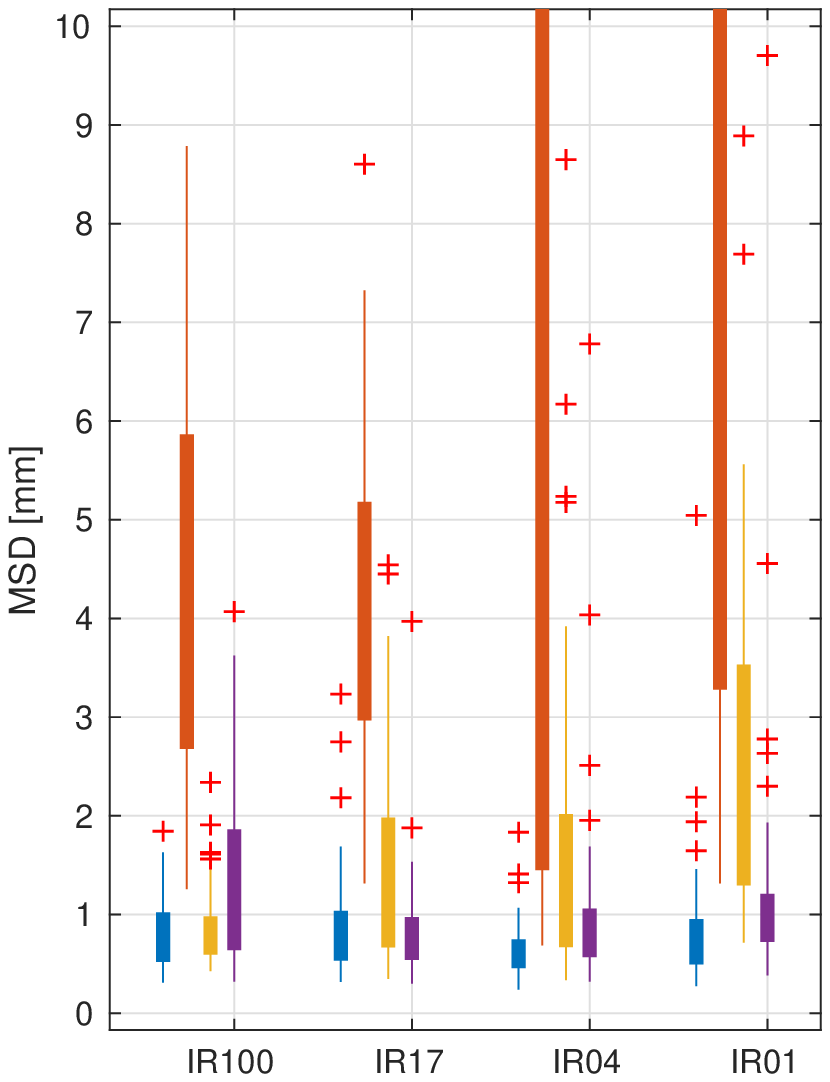}
    \end{subfigure}%
    \begin{subfigure}[b]{0.249\textwidth}
    \centering
        \includegraphics[width=0.99\textwidth, trim= 0cm 0.0cm 0cm 0cm, clip]{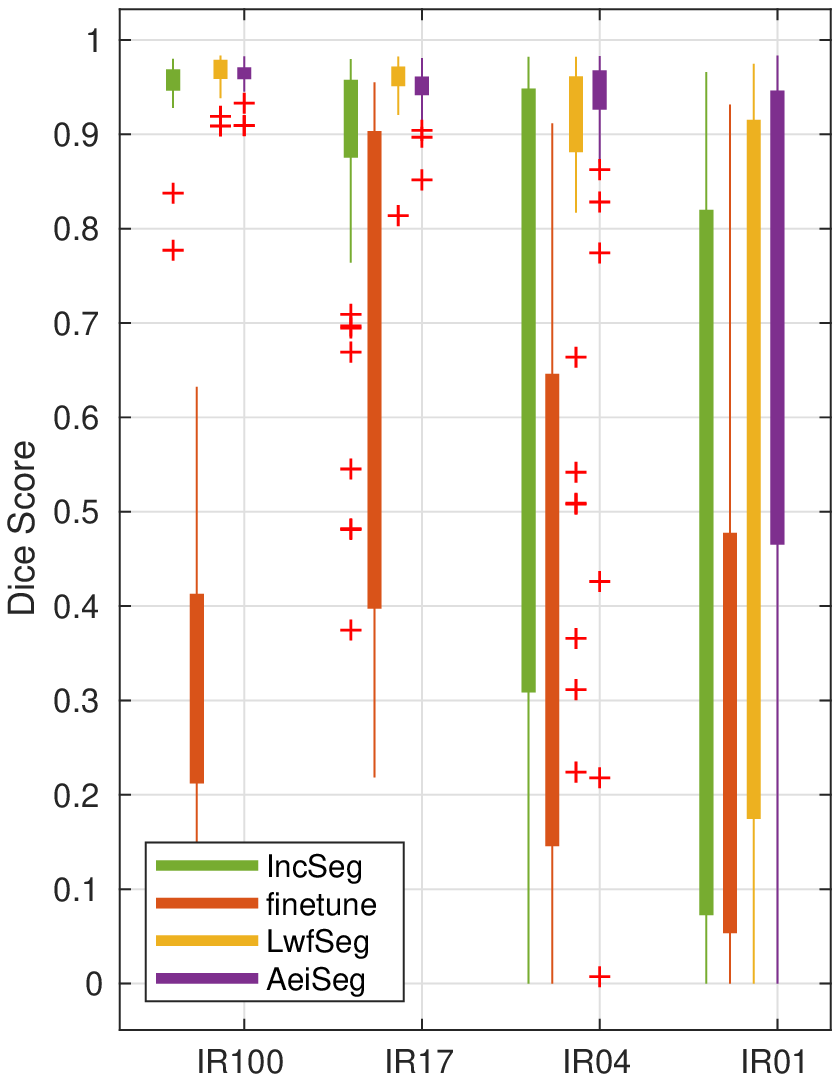}
    \end{subfigure}%
    \begin{subfigure}[b]{0.249\textwidth}
    \centering
        \includegraphics[width=0.99\textwidth, trim= 0cm 0.0cm 0cm 0cm, clip]{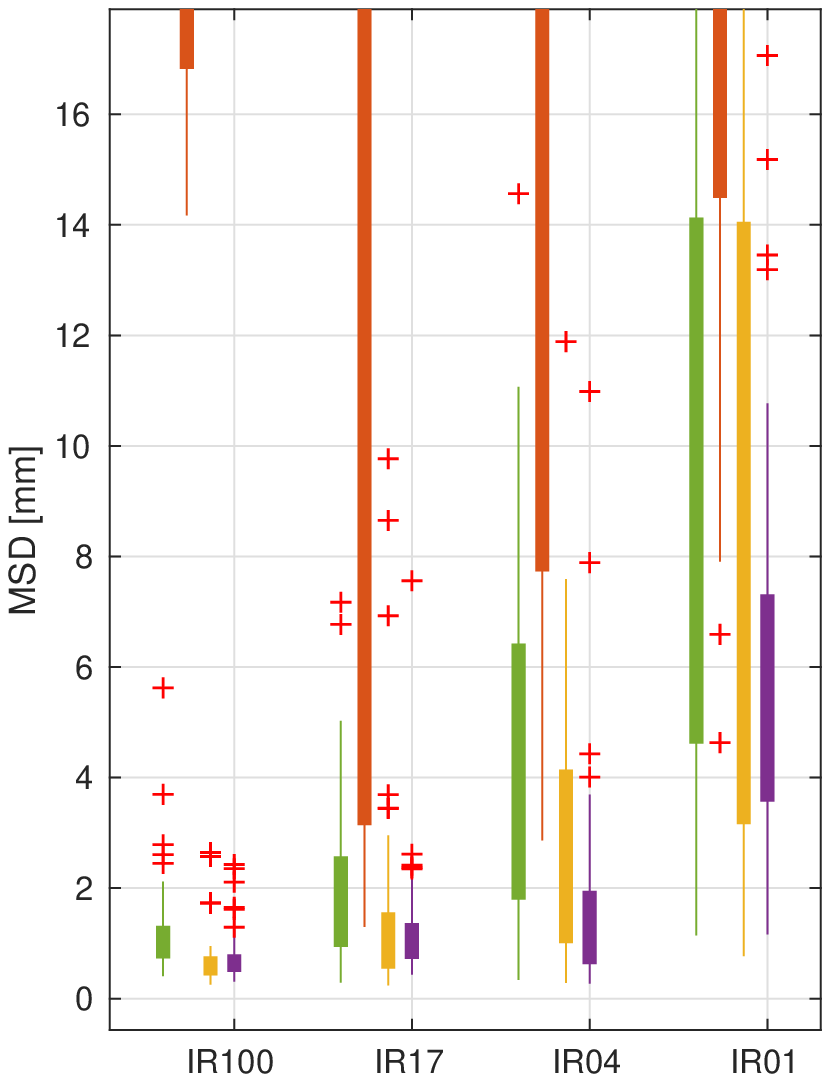}
    \end{subfigure}%
\caption{Comparison between the conventional and incremental methods for 2 holdout sets for the proposed IRs in Table~\ref{tbl:experimental_cases} for Cur$\leftarrow$femur (left) and Inc$\leftarrow$tibia~(right).
Median, outliers, and lower/upper quartiles are indicated.
}
    \label{fig:2holdout_all_methods_BP}
\end{figure*}

In Fig.~\ref{fig:qualitative_comparison}, we show an example slice from the SKI10 dataset along with the segmentation output of the compared incremental methods and the ground truth annotation. 
For quantitative evaluations, we used Dice coefficient score and mean surface distance (MSD) for comparing different methods and IRs.
Fig.~\ref{fig:2holdout_all_methods_BP} shows the distribution of segmentation performances for each test volume over 2 holdout sets, i.e.,\ 50 volumes, where the presented experiments are conducted on two different splits of the dataset (see supplementary material for details of the holdout set generation).
In Table~\ref{tbl:4cases2holdouts}, we show the quantitative results averaged over two holdout sets for all compared methods.
For some method and IR combinations, segmentation output for some images did not contain any foreground labels (i.e.,\ MSD $=\infty$).
We excluded these volumes from the average score calculation in Table~\ref{tbl:4cases2holdouts}, indicating the number of such omitted volumes in parentheses.

\begin{table*}
\centering
\caption{Average Dice and MSD metrics for 2 holdout sets of the experiments in Fig.~\ref{fig:2holdout_all_methods_BP}.
Number of volumes (out of 50) omitted from the average are given in parentheses.
Results superior to conventional methods are shown in bold.
A baseline U-net trained on all data Cur$\cup$Inc as a segmentation upperbound achieves Dice of 94.6\% and 96.2\% for Cur and Inc, respectively.}
\label{tbl:4cases2holdouts}
\adjustbox{max width=\textwidth}{
\begin{tabular}{r|cc|cc|cc|cc}
& \multicolumn{8}{c}{ Dice Score [\%] } \\ \hline
& \multicolumn{2}{c}{ IR100 } \vline & \multicolumn{2}{c}{ IR17 } \vline& \multicolumn{2}{c}{ IR04 } \vline& \multicolumn{2}{c}{ IR01 } \\ \hline
Method & Cur & Inc & Cur & Inc & Cur & Inc & Cur & Inc \\ \hline
CurSeg & 96.0 & - & 95.5 & - & 96.6 & - & 95.2 & -\\
IncSeg & - & 95.3 & - & 87.2 & - & 68.7\,(1) & - & 51.1\,(5)\\ \hline
finetune & 84.0 & 31.4 & 83.7 & 64.4 & 74.1\,(1) & 39.8\,(1) & 57.8\,(1) & 30.8\,(1) \\ \hline
ReSeg & 95.8 & \textbf{95.6} & 89.6 & \textbf{94.2} & 93.4 & \textbf{82.1} & 90.1 & \textbf{57.3} \\ \hline
LwfSeg & 95.6 & \textbf{96.6} & 93.5 & \textbf{95.6} & 88.2 & \textbf{86.5} & 83.7 & \textbf{55.4} \\
AeiSeg & 95.8 & \textbf{96.3} & 95.4 & \textbf{94.8} & 93.7 & \textbf{89.8} & 92.5 & \textbf{70.9\,(1)} \\ \hline
& \multicolumn{8}{c}{ Mean Surface Distance [mm] } \\ \hline
& \multicolumn{2}{c}{ IR100 } \vline& \multicolumn{2}{c}{ IR17 } \vline& \multicolumn{2}{c}{ IR04 } \vline& \multicolumn{2}{c}{ IR01 } \\ \hline
& Cur & Inc & Cur & Inc & Cur & Inc & Cur & Inc \\ \hline
CurSeg & 0.80 & - & 0.90 & - & 0.65 & - & 0.86 & - \\
IncSeg & - & 1.21 & - & 2.04 & - & 4.71\,(1) & - & 9.77\,(5) \\ \hline
finetune & 4.31 & 20.63 & 4.25 & 10.81 & 5.99\,(1) & 16.89\,(1) & 8.27\,(1) & 16.97\,(1) \\ \hline
ReSeg & 1.80 &\textbf{ 0.90} & 4.75 &\textbf{ 1.66} & 1.78 & \textbf{3.39} & 3.19 & \textbf{7.15} \\ \hline
LwfSeg &  0.87 & \textbf{0.69} & 1.45 & \textbf{1.60} & 1.72 & \textbf{2.79} & 3.04 & \textbf{8.45} \\
AeiSeg & 1.29 & \textbf{0.77} & \textbf{0.85} & \textbf{1.26} & 1.06 & \textbf{1.70} & 1.28 & \textbf{5.77\,(1)} \\
\end{tabular}
}
\end{table*}

\begin{figure*}[t]
\centering
	\begin{subfigure}[b]{0.249\textwidth}
    \centering
        \includegraphics[width=0.99\textwidth, trim= 0cm 0.0cm 0cm 0cm, clip]{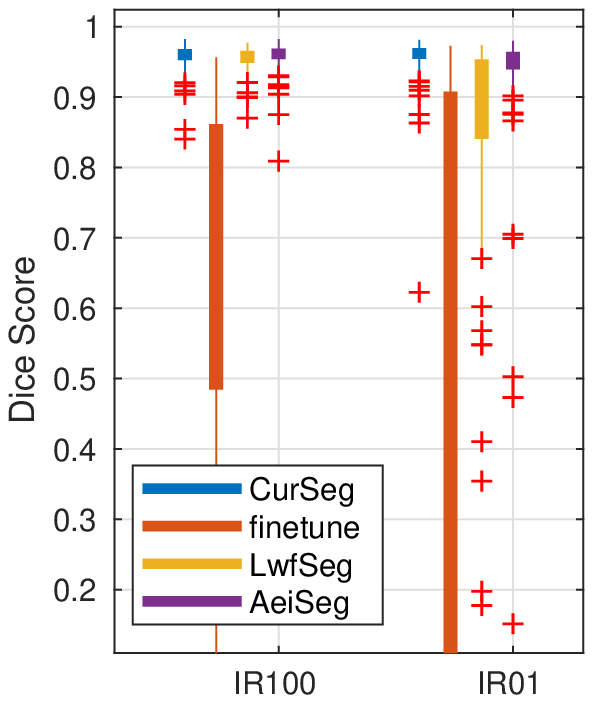}
    \end{subfigure}%
    \begin{subfigure}[b]{0.249\textwidth}
    \centering
        \includegraphics[width=0.99\textwidth, trim= 0cm 0.0cm 0cm 0cm, clip]{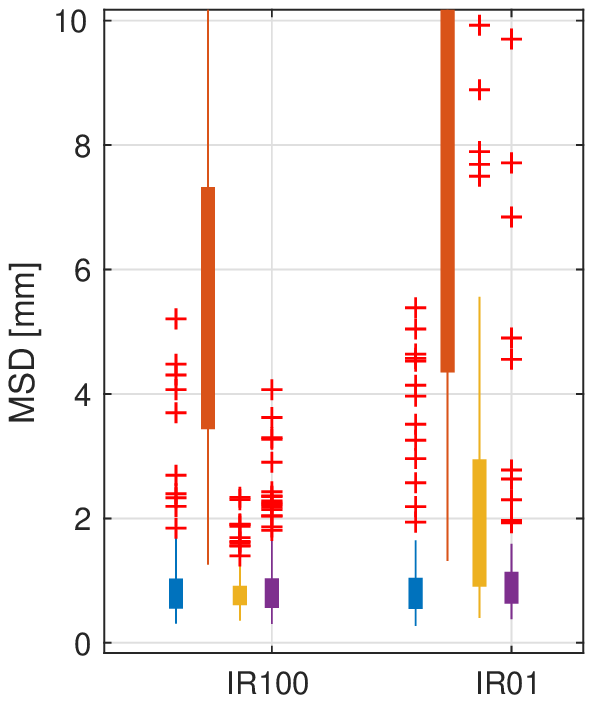}
    \end{subfigure}%
    \begin{subfigure}[b]{0.249\textwidth}
    \centering
        \includegraphics[width=0.99\textwidth, trim= 0cm 0.0cm 0cm 0cm, clip]{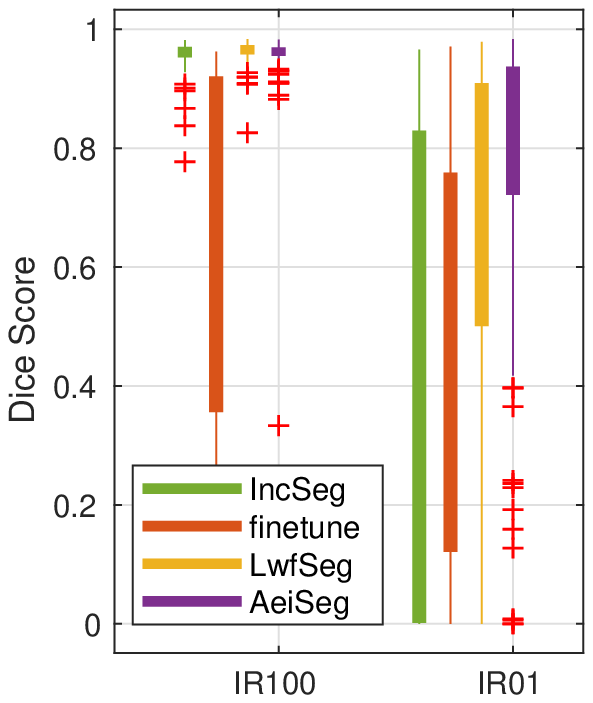}
    \end{subfigure}%
    \begin{subfigure}[b]{0.249\textwidth}
    \centering
        \includegraphics[width=0.99\textwidth, trim= 0cm 0.0cm 0cm 0cm, clip]{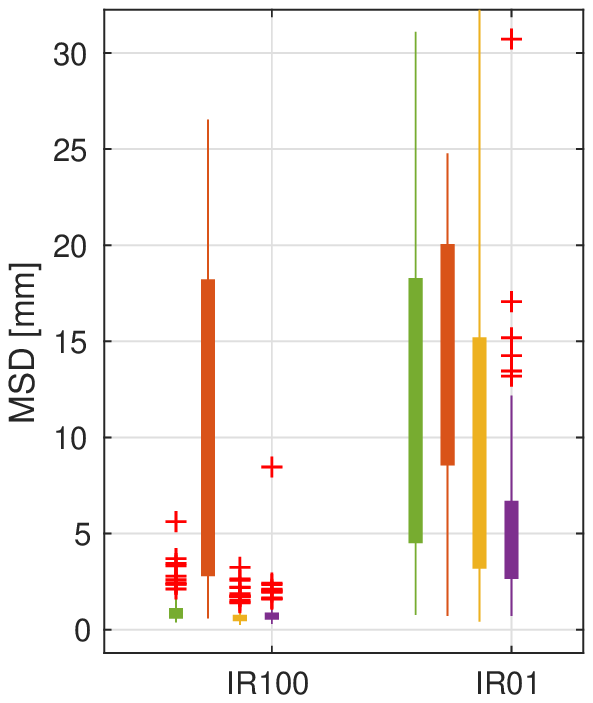}
    \end{subfigure}%
\caption{Comparison between the conventional and incremental methods for 5 holdout sets for 2 extreme IRs for both Cur (left) and Inc (right) classes.}
    \label{fig:5holdouts_BP}
\end{figure*}

To increase the statistical significance of the two extreme IR scenarios (IR100 and IR01), we ran additional 3 holdout experiments each.
We show the distribution of the results from $25\times5=125$ volumes in Fig.~\ref{fig:5holdouts_BP}, where the failed cases can be observed as the outliers in the Dice metric.
The averaged metrics are shown in Table~\ref{tbl:2cases5holdouts}, again omitting completely failed segmentations while indicating their number in parentheses.

\begin{table}
\centering
\caption{Average Dice and MSD 
for 2 extreme IRs for 5 holdout sets in Fig.~\ref{fig:5holdouts_BP}.
Number of omitted volumes (out of 125) are given in parentheses.
Incremental methods superior to conventional approaches are in bold.}
\label{tbl:2cases5holdouts}
\adjustbox{max width=\linewidth}{
\begin{tabular}{r|cc|cc|cc|cc}
& \multicolumn{4}{c}{ Dice Score [\%] } \vline & \multicolumn{4}{c}{ Mean Surface Distance [mm] } \\ \hline
& \multicolumn{2}{c}{ IR100 } \vline & \multicolumn{2}{c}{ IR01 } \vline & \multicolumn{2}{c}{ IR100 } \vline & \multicolumn{2}{c}{ IR01 } \\ \hline
Method & Cur & Inc & Cur & Inc & Cur & Inc & Cur & Inc \\ \hline
CurSeg & 95.7 & - & 95.6 & - & 0.98 & - & 1.06 & - \\
IncSeg & - & 95.7 & - & 57.8\,(25) & - & 1.01 & - & 7.85\,(25) \\ \hline
finetune & 71.8\,(1) & 59.1 & 68.8\,(27) & 49.5\,(13) & 5.77\,(1) & 12.02 & 6.95\,(27) & 12.58\,(13) \\ \hline
LwfSeg & 95.5 & \textbf{96.3} & 86.9 & \textbf{64.7} & \textbf{0.83} & \textbf{0.73} & 2.22 & 9.21 \\
AeiSeg & \textbf{95.8} & 95.5 & 93.3 & \textbf{76.5\,(1)} & 0.99 & \textbf{0.85} & 1.14 & \textbf{5.05\,(1)} \\
\end{tabular}
}
\end{table}

Note that, when learning convolutional kernels, using a fixed physical image resolution may be crucial for capturing anatomical variation in images. 
Our preprocessing step of rescaling images does not preserve such physical dimensions.
Our data augmentation with random scaling may be, to some extent, compensating for this physical dimension loss through scale invariance, but this is a point that has not been well explored in the literature. 
We herein study this comparatively to the original approach above of resized images, in an additional experiment using patches of size $224\times224$\,px  extracted from the original volumes at their isotropic resolution of $0.4$\,mm.
We tested this for the incremental ratios (IR100 \& IR01) for one holdout set, with averaged scores presented in Table~\ref{tbl:resize_vs_patches}. 

To show independence from the incrementing order of structures, we additionally repeated the experiments with current dataset annotations of tibia vs.\ background (Cur*~$\leftarrow$~tibia) and incremental dataset annotations of femur vs.\ background (Inc*~$\leftarrow$~femur), with the results given in Table~\ref{tbl:rv_2cases5holdouts}. 

\section{Discussions}
\label{sec:discussions}

Although the test set from 2 holdout sets were different, it is important to point out that all experimented data IRs had the same test volumes, hence quantitative numbers across different ratios and models for the same class are comparable in Table~\ref{tbl:4cases2holdouts}.
Note that for the incremental class, the baseline incremental method, finetuning, always yields results inferior to the conventional method IncSeg. 
This is simply because during the total of $n_\mathrm{stp}$ steps training, average validation performance of current and incremental classes together does not improve at all after the first few training steps.
Hence, finetuning test results show minimal sign of fitting the Inc dataset.

\begin{table}
\centering
\caption{Comparison of methods when trained using resized images vs patches extracted at original isotropic resolution of $0.4$\,mm, for a single holdout set.
Dice differences over 10\% between resized and patches are marked in bold.
}
\label{tbl:resize_vs_patches}
\adjustbox{max width=\linewidth}{
\begin{tabular}{r|r|cc|cc|cc|cc}
& & \multicolumn{4}{c}{ Dice Score [\%] } \vline & \multicolumn{4}{c}{ Mean Surface Distance [mm] } \tabularnewline  \hline
 &  & \multicolumn{2}{c}{ IR100 } \vline & \multicolumn{2}{c}{ IR01 } \vline & \multicolumn{2}{c}{IR100 } \vline & \multicolumn{2}{c}{ IR01 } \tabularnewline \hline
\multicolumn{2}{c}{ Method } \vline & Cur & Inc & Cur & Inc & Cur & Inc & Cur & Inc \tabularnewline \hline
 \multirow{2}{*}{CurSeg } & Resized & 0.96 & - & 0.97 & - & 0.75 & - & 0.64 & -  \tabularnewline
& Patches & 0.97 & - & 0.97 & - & 1.07 & - & 1.27 & - \tabularnewline
\multirow{2}{*}{ IncSeg } & Resized & - & 0.96 & - & \textbf{0.66} & - & 1.26 & - & 5.80 \tabularnewline
& Patches & - & 0.95 & - & \textbf{0.77} & - & 2.13 & - & 5.16 \tabularnewline \hline
\multirow{2}{*}{ LwfSeg } & Resized & 0.96 & 0.96 & 0.90 & \textbf{0.82} & 0.84 & 0.77 & 2.05 & 3.07 \tabularnewline
& Patches & 0.96 & 0.96 & 0.86 & \textbf{0.71} & 1.77 & 0.90 & 2.93 & 6.53 \tabularnewline
\multirow{2}{*}{ AeiSeg } & Resized & 0.96 & 0.97 & 0.94 & \textbf{0.92} & 0.80 & 0.77 & 1.00 & 4.28 \tabularnewline
& Patches & 0.94 & 0.96 & 0.92 & \textbf{0.78} & 1.33 & 1.45 & 1.74 & 6.79 \tabularnewline
\end{tabular}
}
\end{table}

\begin{table}
\centering
\caption{Average Dice and MSD metrics for 5 holdout sets for two extreme IRs, when the order of incremental structures is switched, i.e.\ Cur*$\leftarrow$tibia and Inc*$\leftarrow$femur.
Number of omitted volumes (out of 125) are in parentheses.}
\label{tbl:rv_2cases5holdouts}
\adjustbox{max width=\linewidth}{
\begin{tabular}{r|cc|cc|cc|cc}
& \multicolumn{4}{c}{ Dice Score [\%] } \vline & \multicolumn{4}{c}{ Mean Surface Distance [mm] } \\ \hline
& \multicolumn{2}{c}{ IR100 } \vline & \multicolumn{2}{c}{ IR01 } \vline & \multicolumn{2}{c}{ IR100 } \vline & \multicolumn{2}{c}{ IR01 } \\ \hline
Method & Cur* & Inc* & Cur* & Inc* & Cur* & Inc* & Cur* & Inc* \\ \hline
CurSeg & 95.2 & - & 95.8 & - & 0.84 & - & 0.79 & - \\
IncSeg & - & 96.1 & - & 57.2\,(10) & - & 0.72 & - & 8.08\,(10) \\ \hline
finetune & 60.3 & 61.6 & 75.2\,(1) & 29.6\,(9) & 10.75 & 9.85 & 4.43\,(1) & 17.73\,(9) \\ \hline
LwfSeg & \textbf{96.1} & 95.3 & 91.4\,(11) & \textbf{60.8\,(1)} & \textbf{0.71} & 0.90 & 1.60\,(11) & 8.17\,(1) \\ 
AeiSeg & \textbf{96.3} & 95.1 & 89.2\,(1) & \textbf{78.5\,(1)} & \textbf{0.66} & 0.95 & 1.62\,(1) & \textbf{4.81\,(1)} \\
\end{tabular}
} 
\end{table}

Both of our proposed class-incremental methods (i.e.,\ LwfSeg and AeiSeg) performed superior to alternatives for all incremental ratios (cf. ~\ref{tbl:4cases2holdouts}).
In addition, considering IncSeg at IR100 as an ideal reference for Inc class segmentation, both LwfSeg and AeiSeg are within 0.5\% of this in Dice even at a low incremental ratio of IR17, and still within 10\% at IR04. 
Furthermore, in Fig.~\ref{fig:2holdout_all_methods_BP}, the results are seen to be more robust with the proposed LwfSeg and AeiSeg methods for samples across all IRs.
Moreover, we also considered a hypothetical upperbound scenario where current and incremental datasets are concurrently available, and trained a conventional (i.e.,\,non-incremental) method, \textit{CurIncSeg} (quantitative results available in the supplementary material). 
CurIncSeg did not outperform our proposed methods LwfSeg and AeiSeg, which supports the arguments in~\cite{hinton2015distilling}, where knowledge distillation helps generalizability and reduces changes of overfitting.

AeiSeg method is seen to tolerate domain shift better than LwfSeg.
Note that as the data imbalance deepens (i.e.\ going from IR100 to IR01), incremental methods (particularly LwfSeg) generally perform worse for the current class Cur: For the tested IRs, Dice score of Lwfseg is, respectively, 0\%, 2\%, 9\%, and 12\% lower compared to CurSeg. 
This phenomenon can be attributed to the fact that SKI10 dataset contains acquisitions from a wide variety of vendors and imaging sequences, where many Inc data, especially at low IRs, may differ substantially from the Cur dataset distribution.
Indeed, without exemplar data, LwfSeg struggles with such substantial domain-shift from Cur to Inc dataset.
A similar trend, despite being less significant is also observed with ReSeg, as it performs inferior to AeiSeg (cf. Table~\ref{tbl:4cases2holdouts}). 
This can be the result of (i) picked exemplars being less useful when Cur is larger or (ii) ReSeg having difficulty in fitting for Inc when soft targets from exemplars are not helpful for the task. 
All the same, ReSeg performs worse than AeiSeg for both metrics and even for the incremental classes.

In Table~\ref{tbl:resize_vs_patches}, the results between resized and patched images are seen to be comparable for Cur dataset, both in balanced and imbalanced scenarios (IR100 \& IR01). 
Nevertheless, when there is an extreme imbalance, i.e.\ low incremental ratio IR, the proposed incremental methods with resized images outperform their patch-based counterparts for Inc data as seen in Table~\ref{tbl:resize_vs_patches}.
This is likely thanks to the resized images preserving the entire field-of-view anatomical context information, which can be readily transferred to the incremental task, becoming a particularly crucial added information given smaller incremental datasets.

We showed the robustness of our methods to the order of structures in Table~\ref{tbl:rv_2cases5holdouts}, where the behavior of methods across different IRs is seen to be invariant to incrementing in the reverse order. 
In this experiment, we retrospectively noticed that the 10 out of 11 fail cases of LwfSeg with IR01 come from the same holdout set, which can be attributed to the single incremental volume being from a substantially different distribution compared to the test set.
Note that without exemplars, knowledge retention depends on the accuracy of the soft targets generated from the new images.
Subsequent to distillation loss using poor soft targets, LwfSeg may indeed \textit{unlearn} the Cur* class, hence the fail cases.
Thanks to retained exemplars, Aeiseg can still successfully segment the Cur* class, achieving a mean Dice score of $93\%$ for the corresponding 10 volumes.
Our methods are presented considering multiple class-incremental learning steps; nevertheless, experimentation studying such multiple incremental setup remains as future work.

Considering the number of convolutional layers in a head, we also experimented with narrower alternatives; i.e.,\ a convolutional layer of kernel size $1\times1$ at the end of the shared body for each dataset.
This has proven to be insufficient for the segmentation task.
Empirically, we found that heads with 2 convolutional layers, each having $n_\mathrm{fil}$ filters, to be satisfactory for the experimented dataset, corroborating our earlier results in~\cite{ozdemir2018learn} on a different dataset of different imaging sequences.

Memory footprint is an essential concern in our framework aiming for lifelong learning.
In our TensorFlow implementation, each new head requires an additional 73\,KB for parameter storage, which is the only additional footprint for LwfSeg.
For AeiSeg, an additional $\leq 21$\,MB was required to store the exemplar set of each class assuming $n_\mathrm{rep}=100$.

We herein focus on a class-incremental scenario. 
For lifelong learning without class-incrementation, i.e.\ with new annotations of an existing class, one possible approach using our setup would be to train the current heads with both $L_\mathrm{seg}$ using the new annotations and $L_\mathrm{dis}$ using the soft targets, in an alternating fashion.
In another scenario, where new annotations can have existing classes as well as new classes, one possible approach would be to increment a new head as proposed in this work. 
Later, during testing, existing class predictions can be fused through logical AND or OR operations between the multiple head predictions, based on the importance of specificity or sensitivity of the task, respectively.
Alternatively, instead of softmax operation, one can apply sigmoid function by treating each channel output as corresponding class versus all classes.

\section{Conclusions}

We have proposed a segmentation framework for lifelong learning with class-incremental annotations.
We have introduced LwfSeg for knowledge distillation to retain segmentation performance of previously trained class(es) when expanding a deep neural network to new class(es).
Thanks to knowledge transfer from pretrained classes through shared body weights, an exciting observation is that our proposed methods may outperform conventional methods for incremental class, especially when there are only a small number of incremental annotations available (few-shot learning).
We also propose an extension, AeiSeg, for successfully maintaining performance of previously trained classes, despite the domain shift caused by the incremental dataset. 
We have shown that, for any incremental dataset ratio, our proposed methods outperform 
the conventional methods that are trained on the incremental classes, indicating the successful knowledge transfer from previous to incremental dataset.
We furthermore observe that for some incremental ratios (e.g., IR17 in Table~\ref{tbl:4cases2holdouts} and IR100 in Table~\ref{tbl:2cases5holdouts}), our methods outrank the segmentations of even the originally-trained model for that class/structure (Cur).
This is indeed an exciting observation, showing that the knowledge gained from the incremental dataset can further improve even the previously-trained class performances.

\section*{Acknowledgment}
This work was funded by the Swiss National Science Foundation and a Highly Specialized Medicine grant (HSM2) of the Canton of Zurich.

\bibliographystyle{ieeetr}
\bibliography{mybibfile} 


\clearpage
\appendix
\normalsize

\subsection{Comparison to Upperbound}

As we mentioned earlier in Sec.~\ref{sec:introduction}, retraining models with the complete dataset whenever new class annotations are procured is not feasible in practice.
However, in order to assess the performance of our proposed methods to a hypothetical upperbound, we introduce CurIncSeg, a conventional (i.e.,\,non-incremental) method which has access to both Cur and Inc datasets.
Having trained on Cur and Inc data simultaneously and being validated on the corresponding holdout validation set, we show in Table~\ref{tbl:upperbound} that this upperbound does not necessarily perform superior to our proposed methods LwfSeg and AeiSeg. 
This phenomenon supports the arguments in~\cite{hinton2015distilling} on generalizability of knowledge distillation, reducing changes of overfitting.
Since incremental ratios (IRs) do not apply to CurIncSeg, we emphasize on scores from LwfSeg and AeiSeg that perform better than CurIncSeg in bold.

\begin{table*}[]
\centering
\caption{Average Dice and MSD for the experiments in Fig.~\ref{fig:2holdout_all_methods_BP}.
Number of omitted volumes (out of 50) are in parentheses.
Incremental method results superior to the upperbound model CurIncSeg are shown in bold.}
\label{tbl:upperbound}
\adjustbox{max width=\textwidth}{
\begin{tabular}{@{}l@{}|@{\,}r@{\,}|@{\,}c@{\,}|@{\,}c@{\,}c@{\,}|@{\,}c@{\,}c@{\,}|@{\,}c@{\,}c@{\,}|@{\,}c@{\,}c}
& & N/A & \multicolumn{2}{c}{IR100}\vline& \multicolumn{2}{c}{ IR17 } \vline& \multicolumn{2}{c}{ IR04 } \vline& \multicolumn{2}{c}{ IR01 } \\ \hline
& & CurIncSeg & LwfSeg & AeiSeg & LwfSeg & AeiSeg & LwfSeg & AeiSeg & LwfSeg & AeiSeg \\ \hline  
\multirow{2}{*}{ Dice [$\%$] } & Cur & 94.6 & \textbf{95.6} & \textbf{95.8} & 93.5 & \textbf{95.4} & 88.2 & 93.7 & 83.7 & 92.5 \\
& Inc & 96.2 & \textbf{96.6} & \textbf{96.3} & 95.6 & 94.8 & 86.5 & 89.8 & 55.4 & 70.9\,(1) \\ \hline 
\multirow{2}{*}{ MSD [mm] } & Cur & 1.17 & \textbf{0.87} & 1.29 & 1.45 & \textbf{0.85} & 1.72 & \textbf{1.06} & 3.04 & 1.28 \\ 
& Inc & 0.86 & \textbf{0.69} & \textbf{0.77} & 1.60 & 1.26 & 2.79 & 1.70 & 8.45 & 5.77\,(1) \\
\end{tabular}
}
\end{table*}

\subsection{Training Duration}
The training time for the experiments presented in Table~\ref{tbl:4cases2holdouts} was 440 GPU hours. 
Additional holdout experiments for 2 IRs in Table~\ref{tbl:2cases5holdouts} took an additional 351 GPU hours. 
Due to the substantial increase in time required for an epoch for patches that cover the full resolution images and with the overlaps between patches, an additional 182 GPU hours was required for the quantitative evaluation presented in Table~\ref{tbl:resize_vs_patches}.
The experiment results shown in Table~\ref{tbl:rv_2cases5holdouts} to show independence from the order of structures required 459 GPU hours.
The training time for all 5 methods; i.e.,\ CurSeg, IncSeg, finetune, LwfSeg, AeiSeg, presented in Tables~\ref{tbl:4cases2holdouts}~and~\ref{tbl:2cases5holdouts} averaged over 14 experiments are 9.0, 9.0, 13.3, 13.8, and 11.4~hours, respectively. 

\subsection{Convergence of Models}
\begin{figure*}[]
\centering
	\begin{subfigure}[b]{0.495\textwidth}
    \centering
        \includegraphics[width=0.95\textwidth, trim= 0cm 0.0cm 0cm 0cm, clip]{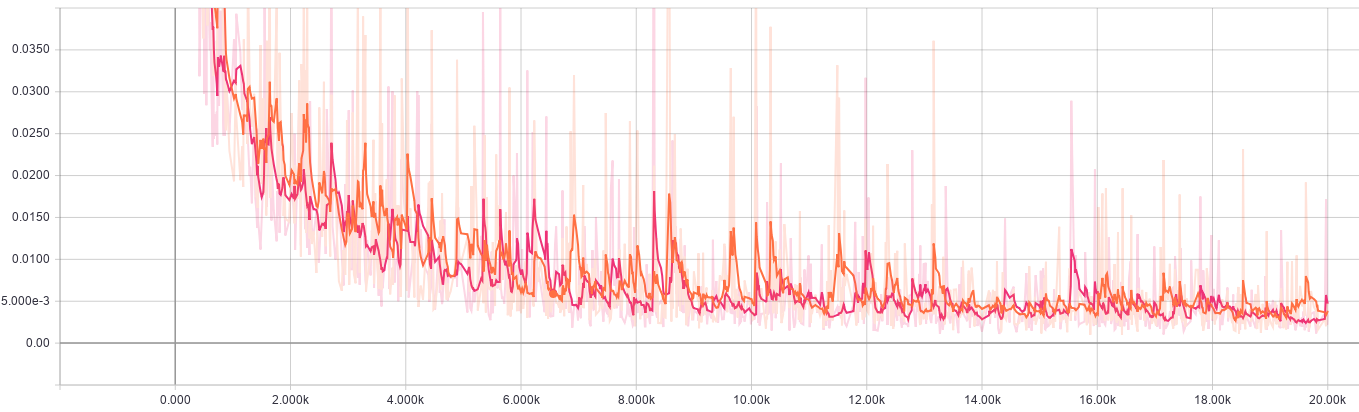}
    \end{subfigure}%
    \begin{subfigure}[b]{0.495\textwidth}
    \centering
        \includegraphics[width=0.95\textwidth, trim= 0cm 0.0cm 0cm 0cm, clip]{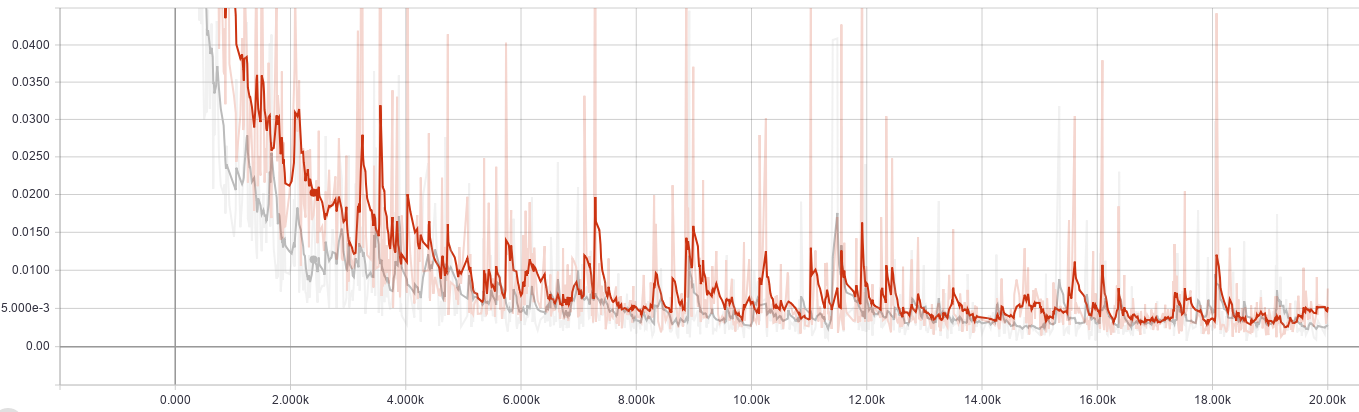}
    \end{subfigure}%
    \\%
    \begin{subfigure}[b]{0.495\textwidth}
    \centering
        \includegraphics[width=0.95\textwidth, trim= 0cm 0.0cm 0cm 0cm, clip]{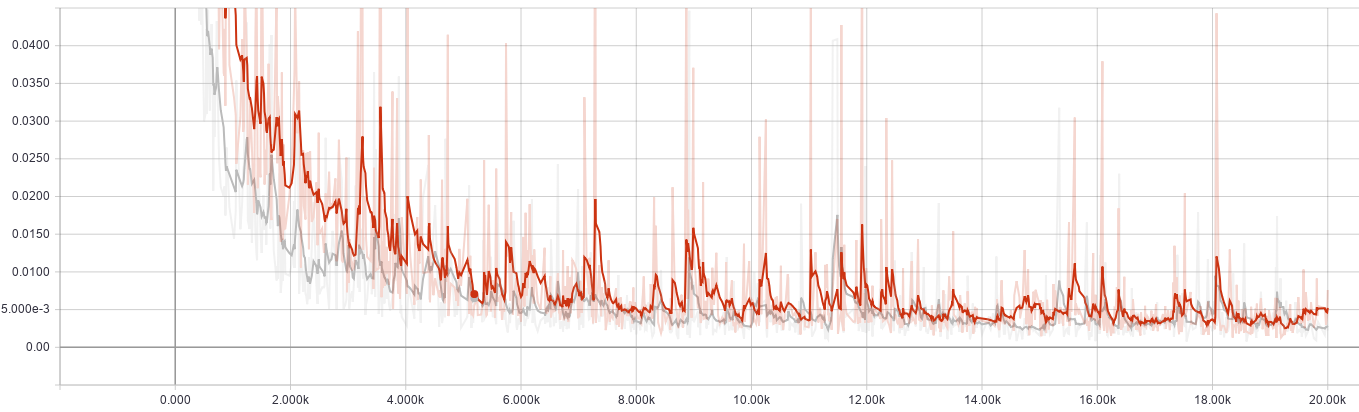}
    \end{subfigure}%
    \begin{subfigure}[b]{0.495\textwidth}
    \centering
        \includegraphics[width=0.95\textwidth, trim= 0cm 0.0cm 0cm 0cm, clip]{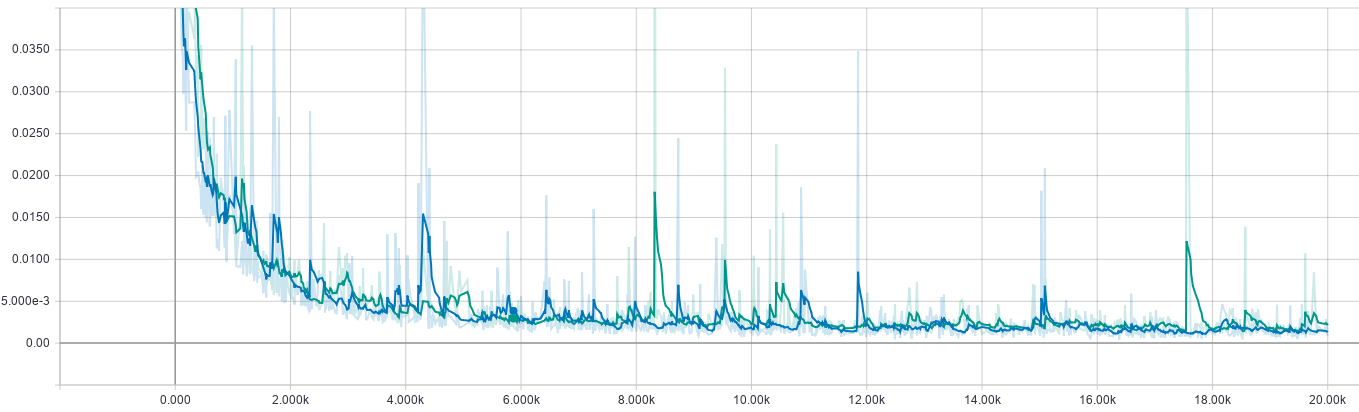}
    \end{subfigure}%
\caption{Training loss for IncSeg across 2 holdout sets (c.f. Table~\ref{tbl:4cases2holdouts}) for the experimented 4 incremental ratios (IR100 top left to IR01 bottom right) throughout $n_\mathrm{stp}=20000$ training iterations.}
    \label{fig:incseg_l_seg}
\end{figure*}

Looking at the Table~\ref{tbl:4cases2holdouts}, one can suggest that IncSeg was surpassed by our incremental methods simply because it was not converged at the end of $n_\mathrm{stp}$ steps. 
Looking at Fig.~\ref{fig:incseg_l_seg}, we were convinced it was not the case for any of the experimented IRs. 

\subsection{Used Software}
All the experiments have been conducted using TensorFlow 1.8.
For computing the evaluation metrics, we have used the \textit{MedPy} Python library.

\subsection{Dataset Evaluation}
In the public SKI10 Challenge dataset, there are 100 annotated knee volumes, each with an identifier (ID) ranging from 1 to 100.
Below, we list the randomly shuffled indices (cf. Listing~\ref{lst:rand_holdout}) used for each holdout set for the purpose of reproducibility. 
From these set of shuffled images, the 4 sets of the current, incremental, validation, and test data are then selected consecutively in the given order, based on Table~\ref{tbl:experimental_cases}.
Therefore, for all incremental settings for a given hold-out experiment, the validation and test sets are exactly the same, allowing us to make direct comparison between IRs of that set.
\begin{itemize} 
\item Holdout set 1:\\
\small
47, 100, 54, 13, 45, 20, 1, 68, 71, 55, 33, 38, 58, 53, 85, 60, 97, 84, 63, 
40, 28, 34, 69, 90, 57, 42, 46, 62, 75, 77, 7, 10, 78, 3, 24, 5, 74, 4, 26, 
15, 31, 29, 87, 92, 44, 16, 73, 88, 21, 56, 94, 89, 41, 51, 93, 9, 17, 99, 
18, 61, 19, 91, 43, 27, 11, 67, 37, 96, 83, 80, 98, 32, 22, 76, 82, 36, 30, 
50, 59, 49, 48, 6, 95, 81, 2, 86, 25, 66, 39, 23, 35, 12, 64, 70, 72, 8, 14, 
52, 79, 65
\normalsize
\item Holdout set 2:\\
\small
18, 70, 7, 8, 58, 40, 21, 46, 50, 19, 54, 64, 52, 36, 85, 3, 24, 94, 72, 67, 57, 37, 23, 96, 68, 32, 83, 55, 66, 26, 48, 11, 6, 98, 38, 31, 14, 27, 13, 4, 81, 74, 80, 20, 5, 89, 47, 33, 22, 9, 78, 45, 65, 97, 63, 56, 43, 1, 87, 84, 86, 35, 34, 2, 93, 15, 77, 53, 100, 60, 79, 42, 39, 71, 62, 59, 75, 16, 41, 69, 95, 82, 10, 44, 28, 92, 25, 17, 76, 99, 88, 12, 30, 73, 91, 49, 51, 90, 61, 29
\normalsize
\item Holdout set 3:\\ \small
91, 98, 23, 57, 3, 77, 36, 24, 30, 82, 44, 99, 85, 79, 19, 48, 81, 87, 8, 56, 40, 83, 29, 47, 42, 18, 89, 45, 11, 39, 43, 74, 20, 52, 92, 53, 50, 69, 62, 66, 58, 12, 59, 54, 6, 84, 38, 60, 5, 15, 86, 14, 25, 78, 90, 68, 27, 33, 10, 32, 17, 100, 88, 9, 96, 76, 67, 41, 80, 97, 93, 65, 70, 49, 73, 37, 95, 46, 4, 1, 63, 2, 13, 64, 71, 55, 7, 22, 61, 16, 35, 34, 72, 31, 28, 94, 75, 26, 51, 21
\normalsize
\item Holdout set 4:\\ \small
12, 82, 42, 95, 54, 5, 3, 60, 4, 33, 89, 24, 21, 65, 35, 43, 17, 91, 50, 61, 2, 59, 81, 7, 34, 53, 87, 80, 62, 38, 32, 27, 77, 67, 79, 52, 100, 49, 98, 20, 22, 84, 48, 94, 85, 96, 9, 41, 29, 44, 39, 92, 31, 75, 66, 69, 45, 70, 99, 18, 46, 6, 97, 26, 83, 37, 11, 16, 88, 90, 68, 36, 57, 19, 8, 74, 93, 14, 64, 56, 55, 78, 23, 10, 63, 13, 47, 1, 15, 86, 40, 72, 30, 28, 76, 25, 51, 58, 71, 73
\normalsize
\item Holdout set 5:\\ \small
84, 54, 71, 46, 45, 40, 23, 81, 11, 1, 19, 31, 74, 34, 91, 5, 77, 78, 13, 32, 56, 89, 27, 43, 70, 16, 41, 97, 10, 73, 12, 48, 86, 29, 94, 6, 67, 66, 36, 17, 50, 35, 8, 96, 28, 20, 82, 26, 63, 14, 25, 4, 18, 39, 9, 79, 7, 65, 37, 90, 57, 100, 55, 44, 51, 68, 47, 69, 62, 98, 80, 42, 59, 49, 99, 58, 76, 33, 95, 60, 64, 85, 38, 30, 2, 53, 22, 3, 24, 88, 92, 75, 87, 83, 21, 61, 72, 15, 93, 52
\normalsize
\end{itemize}

\begin{table*}
\normalsize
\begin{lstlisting}[caption={Generation of different holdout set indices.} ,label={lst:rand_holdout},language=Python]
import numpy as np
def generateHoldoutSet(seed):
    ho_set = np.arange(1,101)
    prng = np.random.RandomState(seed)
    prng.shuffle(ho_set)
    return ho_set
def holdoutSet(setID):
    assert isinstance(setID, int)
    assert (setID>=1 and setID <=5)
    seed_dict = {1:1991, 2:1881, 3:1938, 4:905, 5:42}
    return generateHoldoutSet(seed=seed_dict[setID])
\end{lstlisting}
\end{table*}

\subsection{Architecture}

In Table~\ref{tbl:architecture}, we show the hyperparameters of the architecture we have used in this work. 
For some of the filters, we have assigned a name to the output tensor such that skip connections~\cite{Ronneberger2015unet} can be defined.

\begin{table*}
\centering
\caption{Details of the architecture used in our proposed method. 
Based on the incremental stage of the network, there can be one or more \textit{head(s)} which take the shared body output (o1) as its input.
conv3-32-bn resembles $32$ 2D convolutional filters with a kernel size of $3\times3$ with batch normalization applied before ReLU activation function.
All spatial dropout layers have a drop rate of 0.5. 
All max-pooling and deconvolutional filters have a stride of 2 in each axis.}
\label{tbl:architecture}
\adjustbox{max width=\textwidth}{
\begin{tabular}{cc|cc}
\hline
\multicolumn{2}{c}{ Shared Body } \vline&  \multicolumn{2}{c}{ Head } \\ \hline
Filter & Name & Filter & Name \\ \hline 
input (224 x 224 x 1) & i1 & input (224 x 224 x 32) & o1 \\ 
conv3-32-bn & & conv3-32-bn & \\
conv3-32-bn & c1 &   dropout &   \\  
maxpool &   & conv3-32-bn & \\ 
conv3-64-bn & &   dropout &   \\
conv3-64-bn & c2 & conv1-nc-bn & \\  
maxpool &   &   softmax &   $h_i$ \\
conv3-128-bn & & & \\
conv3-128-bn & c3 & & \\ 
maxpool &   & & \\ 
conv3-256-bn & & & \\
conv3-256-bn & c4 & & \\  
maxpool &   & & \\  
dropout &   & & \\
conv3-512-bn & & & \\
conv3-512-bn & c5 & & \\  
dropout &   & & \\  
deconv4-512-bn &   d4 & & \\  
concat(d4,c4) &   & & \\ 
conv3-256-bn & & & \\
conv3-256-bn & c6 & & \\  
dropout &   & & \\  
deconv4-256-bn &   d3 & & \\  
concat(d3,c3) &   & & \\ 
conv3-128-bn & & & \\
conv3-128-bn & c7 & & \\  
dropout &   & & \\  
deconv4-128-bn &   d2 & & \\  
concat(d2,c2) &   & & \\ 
conv3-64-bn & & & \\
conv3-64-bn & c8 & & \\  
dropout &   & & \\  
deconv4-64-bn &  d1 & & \\  
concat(d1,c1) &   & & \\ 
conv3-32-bn & & & \\
conv3-32-bn & c9 & & \\  
dropout &   o1 & & \\
\end{tabular}
}
\end{table*}

\end{document}